## Author-Based Analysis of Conference versus Journal Publication in Computer Science

Jinseok Kim

Institute for Research on Innovation and Science, University of Michigan
330 Packard Street, Ann Arbor, MI U.S.A. 48104-2910
jinseokk@umich.edu; jinseok.academic@gmail.com

## Abstract

Conference publications in computer science (CS) have attracted scholarly attention due to their unique status as a main research outlet unlike other science fields where journals are dominantly used for communicating research findings. One frequent research question has been how different conference and journal publications are, considering a paper as a unit of analysis. This study takes an author-based approach to analyze publishing patterns of 517,763 scholars who have ever published both in CS conferences and journals for the last 57 years, as recorded in DBLP. The analysis shows that the majority of CS scholars tend to make their scholarly debut, publish more papers, and collaborate with more coauthors in conferences than in journals. Importantly, conference papers seem to serve as a distinct channel of scholarly communication, not a mere preceding step to journal publications: coauthors and title words of authors across conferences and journals tend not to overlap much. This study corroborates findings of previous studies on this topic from a distinctive perspective and suggests that conference authorship in CS calls for more special attention from scholars and administrators outside CS who have focused on journal publications to mine authorship data and evaluate scholarly performance.

Keywords: conference versus journal; conference publication; computer science; authorship pattern; coauthor similarity; title similarity; TF-IDF

## Introduction

A unique characteristic of scholarly communication in computer science (CS hereafter) is the role of conference publications. The CS community regard conference papers as a primary channel of disseminating research outcomes as much as journal papers (Bar-Ilan, 2010; Franceschet, 2010; Glänzel, Schlemmer, Schubert, & Thijs, 2006; Vardi, 2009). Unlike conferences in other science fields, the CS conferences usually attract original research papers, which go through peer-review process. For some conferences, reviews get synthesized by meta-reviewers who are similar to journal editors. The competitiveness and prestige of a conference is often indicated by its acceptance rate. Leading conferences typically show an acceptance rate lower than 20% (Cabanac & Preuss, 2013).

The CS community has discussed its conference-centric publishing culture, especially on the subject of review system and paper quality (Birman & Schneider, 2009; Fortnow, 2009; Ragone, Mirylenka, Casati, & Marchese, 2013). However, the tradition of holding conference publications in high regard has been established as a de facto norm by the practice of computer scientists for decades and has been even legitimized as a formal method of evaluating CS scholars for hiring, promotion, and tenure (Franceschet, 2010; Montesi & Owen, 2008; Vardi, 2009). In addition, large bibliometric databases such as Scopus and Web of Science that had focused on peer-reviewed journal papers began to index conference proceedings for citation counting around mid-2000s (De Sutter & Van Den Oord, 2012).

Accounting for the importance of conference publications in CS, researchers have investigated both conference and journal papers, and sometimes, conference papers alone (e.g., Cavero, Vela, & Caceres, 2014; Franceschet, 2011; González-Albo & Bordons, 2011; Kuhn & Wattenhofer, 2008; Staudt et al., 2012). These bibliometrics studies have provided bird-eye views of authorship characteristics in conferences and journals through advanced data mining techniques applied to large-scale authorship data from, for instance, CiteSeer, Google Scholar, and DBLP.

What is still missing, however, is the understanding of how conference and journal publications are different in terms of individual CS scholars. A few examples of specific questions may include: (1) Does an author who publishes many papers in conferences tend to do so in journals? And (2) how many coauthors of an author work with the author both in conference and journal papers? These questions can be answered when individual authors are considered as a unit of analysis for mining publication data across conferences and journals. Such microscopic observations of publishing patterns per author can be aggregated into insights that help others (esp. from the fields where conferences are not a main venue of communicating research) better understand the distinct publishing culture in CS. A proper understanding of CS publication practice by non-CS people matters because it can guide hiring bodies, funding organizations, and promotion committees "to make more informed decisions" about how to evaluate CS publications (Freyne, Coyle, Smyth, & Cunningham, 2010) as many CS scholars get hired and conduct research in a variety of disciplines. As such, this paper aims to add new knowledge to previous research on the CS authorship by taking an author-based approach to comparing the conference versus journal publications. In the following section, related work is introduced to contextualize this study.

## Related Work

The difference between conference and journal publications has been often discussed with respect to paper quality. For example, Chen and Konstan (2010) found that papers in conferences with low acceptance rate (around 30%) attract comparably the same amount or more citations than journal papers. On the other hand, Freyne et al. (2010) argued that papers in leading conferences show an impact, as measured via the Web of Science's journal citation metrics, similar to that of papers published in intermediate-level ranking journals. Analyzing more than 300,000 publication records, a recent study showed that top CS conference papers are more highly cited than journal papers but papers in medium or low-ranking conferences are not much different in citation frequency from journal papers (Vrettas & Sanderson, 2015). Some researchers, however, raised the concern that citation-based metrics may underestimate the impact of conference papers because bibliometric databases such as ACM's digital library, Scopus, and Web of Science do not fully cover conference publications (De Sutter & Van Den Oord, 2012).

Another branch of research has focused on the extension of conference papers into journal papers or vice versa. In a study of sampled CS scholars, around 25% ~ 33% of CS-related conference papers were found to lead to journal publications (Bar-Ilan, 2010), supporting the similar findings from an interview with 22 editors of 13 CS journals and 122 authors (Montesi & Owen, 2008). Recently, a study surveying 200 papers reported that about 26% of conference papers were extended or republished in journals (Wainer & Valle, 2013). A similar conference-to-journal transition ratio (30%) was reported for research publication in the field of computer vision (Eckmann, Rocha, & Wainer, 2012). These transition ratios are lower than the one for medicine (30~50%) (Miguel-Dasit, Martí-Bonmatí, Sanfeliu, & Aleixandre, 2006) and comparable to or lower than 33% in informetrics (Aleixandre-Benavent, Gonzalez-Alcaide, Miguel-Dasit, Navarro-Molina, & Valderrama-Zurian, 2009). Some researchers have argued that the conference-to-journal transition of research papers can be explained by authors' motivation to enhance research visibility and impact as journal papers are believed to attract more citations than conference papers (González-Albo & Bordons, 2011; Goodrum, McCain, Lawrence, & Giles, 2001; Lisée, Larivière, & Archambault, 2008).

Others have been studying publishing patterns of individual scholars and the structure of collaboration. Their results have shown that, for example, the average number of authors per paper has increased over time across subfields regardless of journals and conferences (Fernandes & Monteiro, 2017). On average conference papers have a larger number of authors (2.69) than journal papers (2.35) (Franceschet, 2011). CS scholars need to seek many coauthors who appear only once in their publication and not ever in others, if they want to publish many papers (Cabanac & Preuss, 2013). Productivity of CS scholars has been shown to increase with the number of subfields in which authors have published journal papers (Subramanyam, 1984). For conferences, scholars who have collaborated with diverse group of scholars are more productive than others (Shi et al., 2011). Finally, the coauthorship network of scholars who appear in conference papers has a smaller average shortest paths than the journal-papers-based network (Franceschet, 2011).

Despite their contributions, the aforementioned studies neglect a relevant research aspect. The extension of conference papers into journal papers has been discussed mostly with regard to topics or contents at a document level, not in terms of individual scholars. Two exceptions exist. The first is the work by Franceschet (2010) comparing the difference of conference versus journal publication counts of three top CS scholar groups (top 10 prolific, top 10 high in $h$-index, and 16 ACM Turing Awardees). The study did not, however, extend the comparison to a larger pool of ordinary CS scholars. Also relevant is a survey of 200 CS papers in Wainer and Valle (2013) which found that among papers extended into subsequent studies, 62% (conference) and 55% (journal) of authors continued to appear in the extended work. The authorship transition was, however, not distinguished for journal-to-conference and conference-to-journal coauthorship transitions per author. In coauthorship network or productivity studies where individual authors are analyzed, only either of conference or journal authorship data are mined without justification of the selection, or, when they are studied together, they are often treated as the same type of publication, not as two different ones. Several digital library services provide a comprehensive authorship profile of individual scholars in CS conferences and journals. However, their fine-grained authorship information for each scholar does not usually result in an aggregated knowledge of how the difference of conference and journal authorship patterns can lead to an overall publishing trend of CS scholars.

Thus, this study aims to complement previous studies by comparing differences of conference versus journal publication patterns at an individual level and understanding the publication trend in CS. For this purpose, especially, this paper calculates ratios of overlapped coauthors and title words per author using a TF-IDF based cosine similarity measure, which is utilized for the first time on this topic. In the present study, the analysis is performed with regard to debut year, publication count, coauthorship, and title keyword. In the next section, the data acquisition and processing for this task is detailed.

## Methodology

### *Data*

Authorship information about CS scholars was obtained from the DBLP computer science bibliography (subsequently referred to as DBLP) (Ley, 2002). Each record has unique publication id, author name(s), year, publication venue, title, etc. DBLP indexes publications in computing research in a broad sense, including major venues in library and information science, and indexes papers published both in conferences and journals. DBLP is highly recognized for its quality control using an algorithmic author name disambiguation supplemented with manual inspection (Ley, 2009; Reitz & Hoffmann, 2013). It has been analyzed in numerous studies for name disambiguation, collaboration mapping, and data management (e.g., Cavero et al., 2014; Franceschet, 2011; Kim & Diesner, 2017; Shi et al., 2011). Recently, the accuracy of DBLP author name disambiguation was evaluated against a labeled dataset of 474 unique scholars in 3,921 publications who have ambiguous surnames such as Li, Kim, Gupta, or Johnson (Kim & Diesner, 2015). The DBLP disambiguation accuracy was 0.952 ($K$-metric) and 0.96 (*Pairwise* F1), which is similar or slightly better than other algorithmic disambiguation techniques (Ferreira, Goncalves, & Laender, 2012).

The XML format of DBLP collection (September 2017 version) was downloaded and parsed using Java parsers provided by DBLP[1]. A total of 3,404,499 conference or journal paper records were selected for analysis after several filtering steps. (1) Publications other than conference and journal papers (such as books, reviews, and thesis) and conference or journal papers without author names, titles, or publication years were excluded. (2) Conference publications in DBLP appeared first in 1959, while the record of journal publications go back to 1936. For the purpose of comparing authorship difference in conferences and journals, papers published before 1959 were omitted. Also, papers published in 2017 were excluded because records for that period are incomplete due to, e.g., the lag time in publisher indexing. (3) Following Cabanac, Hubert, and Milard (2015), papers in CoRR[2] or IACR Cryptology ePrint Archive[3] were excluded. Although they are categorized as journal papers in DBLP, they are not peer-reviewed, their status (e.g., draft, pre-print, or published ones) is unclear, and they often lead to duplicate records (e.g., both a pre-print in CoRR and its journal version paper are recorded in DBLP). (4) Papers that have common titles such as editorial, news, and introduction, *and* appear three or more times in DBLP were filtered. (5) Finally, papers were not included if they have the same titles *and* authors in the *same* venues.

---

[1] Downloaded at dblp.org/xml/release/dblp-2017-09-03.xml.gz

As this study aims to analyze how an author's publishing pattern in conferences and journals is different, only authors who have ever published *both* in conferences and journals were selected for the target population, resulting in a total of 517,763 unique authors. A note is that in DBLP unique authors are represented by name strings. Some authors share names (homonyms) and, if not properly disambiguated, can be mistaken as the same author. To handle these homonymous cases, DBLP team uses a network-based community detection technique as well as manual inspection and assigns four-digit numbers to each distinct authors with the same names (e.g., Wei Wang, Wei Wang 0001, Wei Wang 0002, etc.) (Momeni & Mayr, 2016). In contrast, some unique authors are recorded by two or more name strings (synonyms) in the DBLP raw data. For these cases, the DBLP online service matches different author name strings believed to refer to the same scholar and list them on the scholar's publication profile ("a.k.a" section). For this study, such synonyms were consolidated using the "a.k.a" information[4]. Next, each unique author was assigned a list of her/his conference and journal publications. This process produced a total of 7,652,228 author-publication instances. For example, if author A has published 12 papers in conferences and 8 in journals, then s/he comes to have a list of 20 author-publication instances. Each instance was formatted as follows: author name, venue type (conference or journal), publication year, coauthor names, and paper title. This list was used to measure the differences of conference versus journal publications per author and the outcomes of all authors were aggregated for calculating mean, median, and standard deviation (*SD*) values.

*Measurements*

*Debut Year/Career Year*: A debut year, as a proxy of an academic age, is the year where an author's first publication appears in DBLP. A debut year (i.e. the first publication year) was found to be the strongest predictor of actual age (in terms of birth and PhD years) of scholars (Nane, Lariviere, & Costas, 2017). A limitation is that the debut year of a scholar includes the dormant period of scholarly publication after the last paper has been published. Meanwhile, a career year of a scholar is the duration of publication activity (Milojević, 2012), which is measured as the difference of the last and the first year when her/his publication appears in conferences (i.e., conference career year) or journals (i.e., journal career year).

*Production*: An author's production is the number of publications in DBLP attributed to the author. This is the total frequency of an author's name in the data. Conference and journal paper counts are considered separately for analysis. This counting assigns a full publication credit to an author regardless of the number of coauthors in a paper or the author's rank in the byline. In order to check the relative dominance of conference or journal publication per author, a *PubRatio* is calculated as follows.

---

[2] Computing Research Repository (http://arxiv.org/corr/about)
[3] https://eprint.iacr.org/
[4] The list of 39,152 author name pairs in synonym relation was kindly provided by Florian Reitz at DBLP

$$PubRatio = \frac{(No.\,of\,conference\,publication\,-\,No.\,of\,journal\,publication)}{(No.\,of\,conference\,publication\,+\,No.\,of\,journal\,publication)} \qquad (1)$$

The value varies between -1 (complete dominance of journal publication), 0 (balance), and 1(complete dominance of conference publication).

*Coauthorship*: First, unique coauthors of an individual author are recorded for conferences and journals. Here, the frequency of collaboration is ignored: i.e., only the existence of coauthoring between a pair of an author and her/his coauthor matters. For each author, three coauthor lists are generated: two lists of unique coauthors of an author in conferences and journals, respectively, and the overlap of both lists. The latter is used to calculate *CoauOverlap*, i.e., the ratio of how many coauthors in conferences and journals per author overlap against all unique coauthors.

$$CoauOverlap = \frac{Coauthors\,in\,conference\,\cap\,Coauthors\,in\,journal}{Coauthors\,in\,conference\,\cup\,Coauthors\,in\,journal} \qquad (2)$$

This calculation is a variation of Jaccard Coefficient and its interpretation is intuitive. The *CoauOverlap* varies between 0 (no shared coauthor) and 1 (every coauthor appears both in conference and journal papers).

Second, two coauthor lists (one for conferences and the other for journals) are compared to decide how they are similar with regard to each coauthor's frequency of collaboration with the author. For this, a cosine similarity of coauthor lists, *CoauCosine*, is calculated as follows. Cosine similarity is often used in information retrieval to compare text documents. This study extends its usage to the measurement of coauthor list similarity, following several author name disambiguation studies (e.g., Levin, Krawczyk, Bethard, & Jurafsky, 2012).

$$CoauCosine = \frac{\sum_{i=1}^{n} CC_i \times JC_i}{\sqrt{\sum_{i=1}^{n} CC_i^2} \times \sqrt{\sum_{i=1}^{n} JC_i^2}} \qquad (3)$$

Here, each coauthor list ($CC$ = conference coauthor list and $JC$ = journal coauthor list) is represented as a vector of coauthors where the value of each coauthor ($CC_i$ or $JC_i$) is the TF-IDF weight of her/his appearance. The TF (Term Frequency) counts how often a coauthor has collaborated in each list with a target author (for whom the cosine similarity is calculated), which is normalized by the number of unique coauthors in each list. This normalized TF discounts two lists' similarity when one long list includes all coauthors in the other and they are regarded to be highly similar. The IDF (Inverse Document Frequency) considers how often a coauthor in each list appears in other target authors' coauthor lists, discounting the effect of common coauthors who would make lots of lists to appear similar to each other. It is calculated by counting the total of coauthor lists (regardless of whether they are conference or journal lists) in data, dividing it by the number of coauthor lists containing the specific coauthor, and then getting the logarithm (base 10 in this study) of the output. Finally, the TF-IDF is the product of TF and IDF.

The value of *CoauCosine* varies between 0 (quite dissimilar) and 1 (quite similar) but, unlike *CoauOverlap*, its interpretation may not be straightforward. This can be illustrated in Table 1.

| Coauthors | Case 1 | | | | Case 2 | | | | Case 3 | | | |
|---|---|---|---|---|---|---|---|---|---|---|---|---|
| | A | B | C | D | A | B | C | D | A | B | C | D |
| List 1 | 5 | 3 | 2 | 1 | 5 | 3 | 2 | 1 | 5 | 3 | 2 | 1 |
| List 2 | 20 | 3 | 2 | 1 | 1 | 2 | 3 | 5 | 20 | 12 | 8 | 4 |
| *CoauCosine* | 0.90 | | | | 0.56 | | | | 1.00 | | | |

Let's assume that an author has two coauthor lists and both of them contain A, B, C, and D coauthors. Numbers in each cell represent the TF-IDF weights of each coauthor. As the author have the same sets of coauthors, the *CoauOverlap* values in three cases are the same (= 1.00). Depending on the TF-IDF values, however, the *CoauCosine* values can differ much. In Case 1, A has different weights across lists while other coauthors have constant weights. In Case 2, weights in one list are reversed in order in the other list. In Case 3, weights in one list increase or decrease by the same proportion (i.e., ×4) in the other list. These examples show that *CoauCosine* measures coauthor list similarity in a different way from *CoauOverlap*. Specifically, given a set of overlapping coauthors between conference and journal papers, *CoauCosine* will be high if coauthors collaborate with a target author by a similar order of collaboration frequency both in conference and journal papers (e.g., top frequent coauthors in conference and journal papers are the same, and the next frequent coauthors are the same, etc.), while each coauthor's effect on similarity will be discounted by the length of the list containing the coauthor and the frequency of the coauthor's appearance in other lists.

*Title words*: Two sets of title words in conferences and journals per author are compared to measure the semantic similarity between conference and journal authorship of an author. For this, title words are stop-listed to filter common English words[5] and stemmed with the rule-based Porter algorithm (Porter, 1980)[6]. The *WordOverlap*, i.e., the ratio of unique title words appearing both in conferences and journals against all unique title words per author, is calculated by Jaccard coefficient, as described in equation (4).

$$WordOverlap = \frac{Title\ words\ in\ conference\ \cap\ Title\ words\ in\ journal}{Title\ words\ in\ conference\ \cup\ Title\ words\ in\ journal} \tag{4}$$

The value varies between 0 (no shared word) and 1 (every word appears both in conference and journal papers).

In the same way the coauthor similarity is calculated above, two title word lists (one for conferences and the other for journals) are compared to decide how they are similar with regard to each word's frequency. A cosine similarity of title word lists, *WordCosine*, is calculated as follows.

$$WordCosine = \frac{\sum_{i=1}^{n} CW_i \times JW_i}{\sqrt{\sum_{i=1}^{n} CW_i^2} \times \sqrt{\sum_{i=1}^{n} JW_i^2}} \tag{5}$$

Here, each word list ($CW$ = conference word list and $JW$ = journal word list) is represented as a vector of words where the value of each word ($CW_i$ and $JW_i$) is the TF-IDF weight of its appearance. The TF (Term

---

[5] https://github.com/stanfordnlp/CoreNLP/blob/master/data/edu/stanford/nlp/patterns/surface/stopwords.txt
[6] https://tartarus.org/martin/PorterStemmer/

Frequency) counts how often a word has appeared in each list for a target author, which is normalized by the number of unique words in each list. The IDF (Inverse Document Frequency) considers how often a word in each list appears in other target authors' word lists. It is calculated by counting the total of word lists (regardless of whether they are conference or journal lists) in data, dividing it by the number of word lists containing the specific word, and then getting the logarithm (base 10 in this study) of the output.

The value of *WordCosine* varies between 0 (quite dissimilar) and 1 (quite similar). Given a set of overlap title words between conference and journal papers, *WordCosine* will be high if words appear by a similar order of frequency both in conference and journal papers (e.g., top frequent words in conference and journal papers are the same, and the next frequent words are the same, etc.), while each word's effect on similarity will be penalized by the length of the list containing it and its frequency in other lists.

Analysis

*Debut Year/Career Year*

An author's academic debut, as a proxy of academic age, is the first year when a publication written by the author appears in conferences and/or journals. As the submission-to-publication time in conferences is shorter than that in journals (Birman & Schneider, 2009; Fortnow, 2009; Freyne et al., 2010), it may be inappropriate to directly compare publication years of conferences and journals to find which type of venue, journal or conference, serves the debut stage of an author. In this study, however, publication year is considered as it is. Out of 517,763 authors who have ever been active in both conferences and journals during 1959~2016 period, 64.20% (332,394) of them first published at a conference and 25.44% (131,707) in a journal. A total of 53,663 authors (10.36%) made a debut on the same year both in a conference and a journal. Thus, for CS scholars, conferences are the main debut venue.

In Figure 1, the number of authors per debut year is plotted in three lines: conference-first (solid), journal-first (dotted), and simultaneous (double). The figure shows that CS scholars made more first debuts as authors in conferences than in journals starting from early 1980s (see the inset figure) and the gap between conference and journal has increased ever since. This observation is in line with the statement that conference-based publication began to dominate the CS research since the early 1980s (Vardi, 2009).

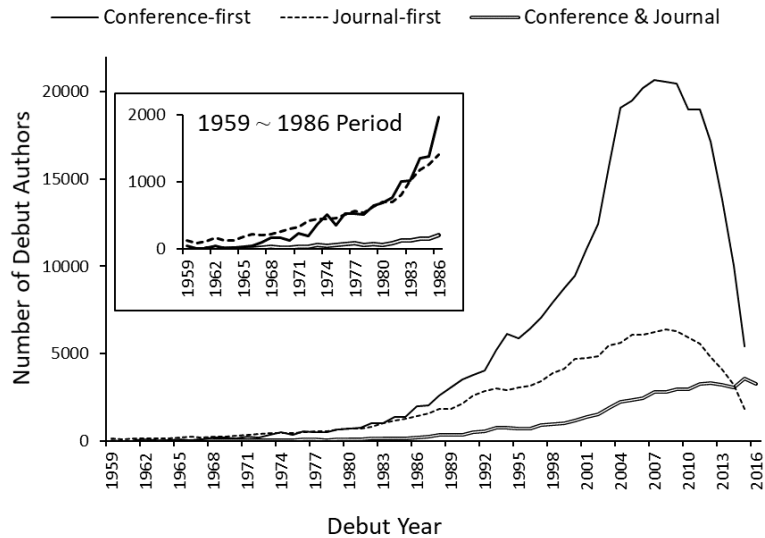

**Figure 1: Number of Debut Authors Per Venue Type for all Years (Main) and the 1959~1986 Period (Inset)**

The figure also tells that both conference-first and journal-first debuts made their peak around 2007~2008, while the both-conference-journal debut is consistently increasing. This observation should be, however, taken with caution because this study only considers authors who have published at least once both in conferences and journals as of 2016. Authors who have published in conferences or journals in recent years but not in the other outlets as of 2016 are not detected in this study. Such lack of data coverage might lead to the abrupt decline of conference-first and journal-first debut trends after 2010 in the figure. This explanation is supported by the observation that authors who had first appeared in conferences made their first appearance in journals on average 3.96 years later ($SD$ = 3.79) and those who had first appeared in journals made their debut in conferences after on average 5.55 years ($SD$ = 5.75).

An author's academic career is the length between the first and last years of publication. The mean academic career of CS scholars, based on both conference and journal papers, is 10.08 year. Conference career (median = 4; mean = 6.96) lasts slightly longer than journal career (median = 3; mean = 5.66). Large numbers of authors have only one career year in each outlet: conferences (161,647) and journals (210,496). One career year means that an author publishes only one paper in conferences and/or journals. This indicates that many authors appear only once in conferences and/or journals and have not appeared again until 2016. This might be due to the fact that many of CS students who have coauthored with academic advisors might go to industry after graduation, or that many scholars made their debut in recent years and have not yet published their next papers. Especially, the larger number of one-time publishing authors in journals than in conferences may be related to the conjecture that journal papers involve higher costs in terms of time and efforts than conference papers (Bar-Ilan, 2010; Montesi & Owen, 2008).

Does an author who has a long conference career tend to have also a long journal career or vice versa? The association between conference and journal careers was tested through Kendall's rank-order correlation (tau, $\tau$) because career distribution is highly skewed and has many tied values. The test showed an intermediate level of correlation ($\tau$ = 0.42). This indicates that career years in conferences and journals per author is not necessarily proportional, implying some authors have a preference toward conferences or journals.

*Production*

On average, CS scholars have published 14.78 papers: 9.12 papers in conferences and 5.65 papers in journals. This indicates that conferences are a more prevalent channel of research communication. Authors who publish many papers in conferences (or journals) tend to publish many papers in journals (or conferences), but the strength of association is weak ($\tau = 0.42$).

Production distribution of authors in conferences, journals, and both, can be plotted on a cumulative log-log plot to see its skewedness. In Figure 2, the horizontal axis represents the number of papers ($x$) and the vertical axis the ratio of authors who have written $x$ or more publications over the total number of authors. All the plots show that a small group of authors have produced many papers while most authors have published a few. Especially, the plots were fitted to power law slopes for 90% of authors in conferences (circles in Figure 2, $1 \leq x \leq 21$, $y = 1.32x^{-1.80}$, $R^2 = 0.98$) and journals (triangles in Figure 2, $1 \leq x \leq 13$, $y = 1.15x^{-1.92}$, $R^2 = 0.99$), implying that the production of CS scholars in both conferences and journals may be modeled to follow predictable patterns (Lotka, 1926).

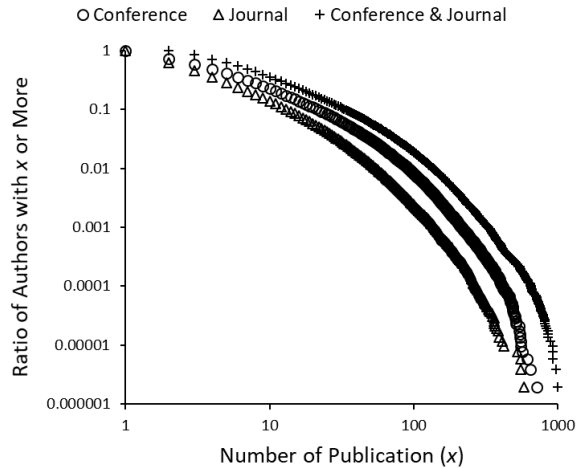

**Figure 2: A Cumulative Log-Log Plot of Production Distribution**

A degree of conference-journal publication balance per author can be assessed by *PubRatio*. About a half (281,371; 54.34%) of scholars have more publications in conferences than in journals (i.e., *PubRatio* > 0), while slightly more than one fifth (123,320; 23.82%) of scholars have published more often in journals (i.e., *PubRatio* < 0). Among those (113,072; 21.84%) who have published equally in conferences and journals (i.e., *PubRatio* = 0), almost two thirds (82,830/113,072; 73.25%) have only two publications; i.e., each in one conference and one journal.

The mean *PubRatio* of all authors is 0.15, which indicates that CS authors have a tendency to publish a little more in conferences than in journals. The mean *PubRatio* of authors per debut year is plotted in the left sub-figure of Figure 3, showing that the younger CS scholars are, the higher their mean *PubRatio* is. This means that, on average, young computer scientists tend to depend more on conferences than journals for scholarly communication than their older peers. The right sub-figure shows that as CS scholars publish more papers, they tend to publish more in conferences: the mean *PubRatio* starts at zero for two publications, rises towards 0.2 around 10 publications, and then keeps hovering higher than 0.2 until reaching 100 publications. A note is that in order to reduce the noise in visualization hereafter, authors who have published 100 or more papers (9,909 out of 517,763; 1.91%) are aggregated together for the calculation of mean, median, and SD values.

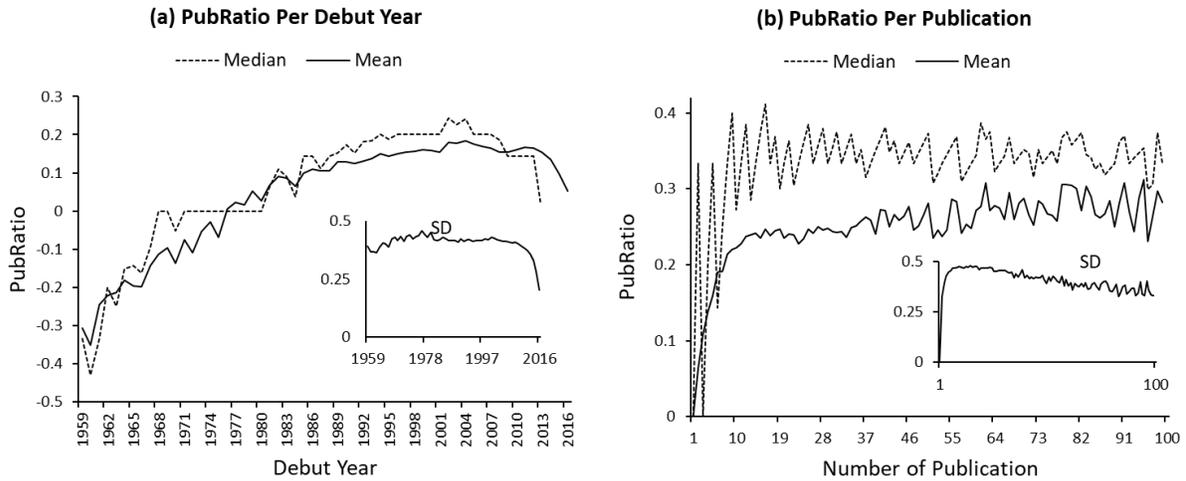

Figure 3: Trends of *PubRatio* per (a) Debut Year and (b) Number of Publication

In the left sub-figure, the mean trend's decline around 2012 may be due to incomplete data. As detailed for Figure 1 in *Debut/Career Year*, authors who have published only in either conferences or journals are excluded from analysis as this study considers only authors publishing both in conferences and journals. For CS scholars who made their debut in recent years, this selection can result in the over-representation of authors who made their debut in a conference and a journal at the same time and have not yet published more (their *PubRatio* is zero). In addition, as shown in the sub-figure (b), authors who have small number of publications tend to have a low *PubRatio*. These factors seem to contribute to the declining *PubRatio* trend for recently debuted authors.

With regard to the conference versus journal preference of CS scholars, an interesting question would be how the debut venue type is associated with an author's choice of publication venue type afterwards. For example, does an author who has first published a paper in a journal tends to prefer journals as s/he continues to publish? For this, the mean *PubRatio* of authors can be calculated per their debut venue type (conference-first, journal-first, or both-conference-journal) over numbers of published papers. The results are shown in Figure 4 with Median and SD trends.

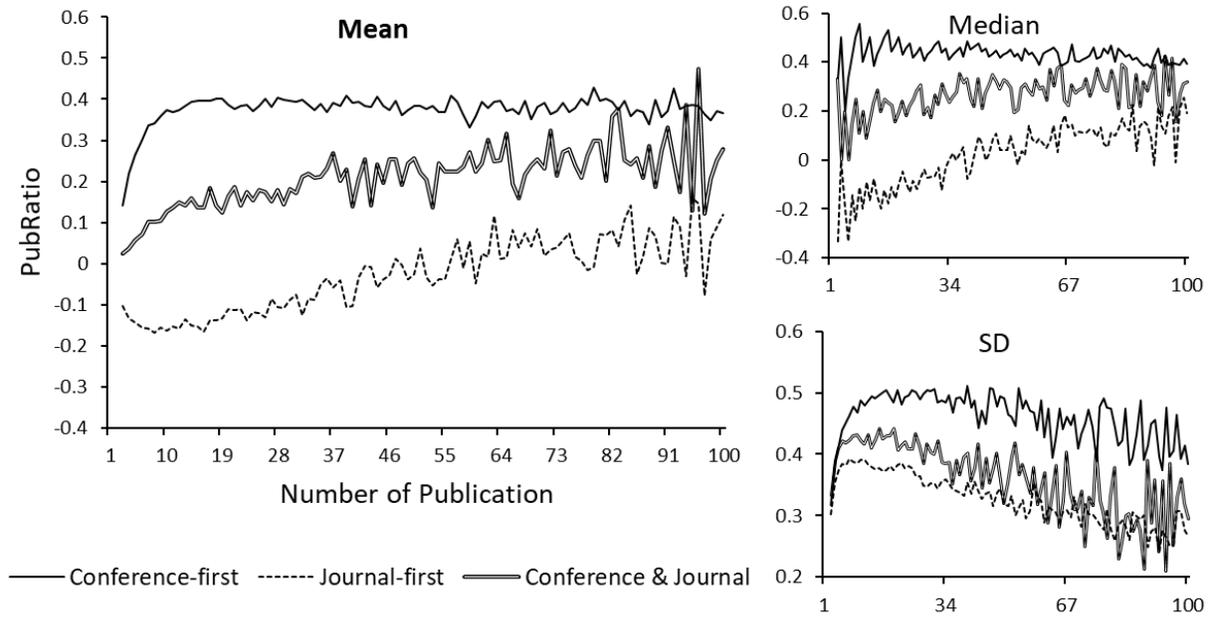

The mean trend of *PubRatio* (left in Figure 4) shows that authors who made their debut in a conference or both in a conference and a journal are likely to prefer conferences to journals as they publish more papers. Their conference-preferred trends are stable (solid line) or slightly increasing (double line). In contrast, journal-debut authors tend to choose journals over conferences in their early publications but keep increasing the ratio of conference papers as they publish more. Their *PubRatio* trend (dotted line) goes up and down around -0.1 until approximately 40 publications and then fluctuates toward/around the zero. Median also shows similar trends. All observations in this section indicate that conferences will continue to be a dominant venue type for CS scholars, if the current trends continue.

*Coauthorship*

The number of authors who have published all papers in journals and conferences as a single author is 2,849 out of 517,763 (0.55%). This means that collaboration is a typical mode of knowledge production for CS scholars. Overall, authors in the dataset have on average 20.08 unique collaborators: 13.93 in conferences and 10.56 in journals. Specifically, 54.44% (281,855/517,763) of CS authors collaborate with a larger number of unique authors (mean 20.33) on conference papers than journal papers (mean 9.70). In contrast, 31.86% (164,966/517,763) of authors have more unique collaborators (mean 14.89) in journals than in conferences (mean 7.33). The remaining 13.70% authors (70,942/517,763) have equal numbers of unique collaborators (mean 3.88) in both outlets.

Does an author who has many collaborators in conferences tend to have many collaborators in journals? A correlation test shows a Kendall's tau of 0.44: an intermediate level of association. How about collaboration across conferences and journals? In other words, do coauthors of a scholar in conferences also tend to appear in journals or vice versa? The overlap in unique coauthors (i.e., coauthors who work together both in conferences and journals with a target author) is on average 4.41 coauthors. For more details, the ratio of coauthor overlap over the total of unique coauthors per author (*CoauOverlap*) is calculated. The mean *CoauOverlap* is 0.29 for all authors in the data. This means that on average 29% of coauthors per author participate both in writing conference and journal papers with the author. In terms of conference collaboration, the overlapping coauthors constitute on average 39.81% of all conference coauthors per author, which means that about 40% of coauthors who ever collaborate in conference papers with an author also appear in the author's journal publications. In terms of journal collaboration, on average 49.52% of coauthors who work together in journal papers with an author also appear in her/his conference publications. These observations can be compared to the paper-level findings in Wainer and Valle (2013) that 62% of authors in conference papers and 55% of authors in journal papers also appeared in extended papers.

In Figure 5 (a), the *CoauOverlap* ratio trend is shown over the number of publication. Roughly until 20 publications, the mean ratio of overlapping coauthors tends to decrease. Although it increases steadily as scholars publish more paper, the mean ratio trend continues to move below 0.3. This implies that CS scholars have a distinct set of collaborators for conferences versus journals. In other words, the collaboration in conference papers does not necessarily lead to that in journal publication or vice versa. Especially, if we assume that coauthors from a conference paper would work together for developing their paper into a journal submission, the mean proportion of overlapping coauthors for conferences (i.e., 39.81%) indicate that many conference papers may not be extended into journal papers, possibly supporting the findings of previous studies (Bar-Ilan, 2010; Vardi, 2009; Wainer & Valle, 2013).

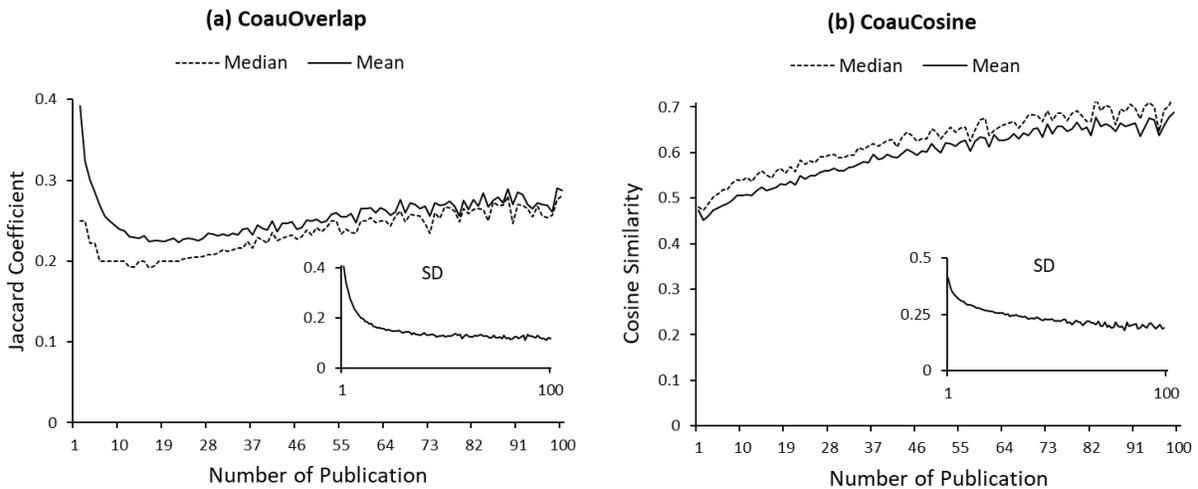

**Figure 5: Trends of (a) Coauthor Overlap and (b) Cosine Similarity per Number of Publication**

With regard to overlapping coauthors, their contributions for authors seem to differ across conferences and journals. According to Figure 5 (b), the cosine similarity of coauthors participating in both conferences and journals starts around 0.47 and keeps increasing up to 0.69. The overall *CoauCosine* values imply that coauthors of an author contribute to conference and journal papers in different ways. For example, coauthors who team up frequently in conferences with an author may collaborate less in journals. In addition, the rising trend of *CoauOverlap* (Figure 5 (a)) and *CoauCosine* (Figure 5 (b)) indicate that as scholars publish more papers, they become to involve more coauthors both in conference and journal papers and to collaborate more frequently with specific coauthors.

The *CoauOverlap* and *CoauCosine* ratio trends were also plotted per debut year in Figure 6. Authors who have published their first publication in recent years (i.e., have short academic ages) tend to show higher *CoauOverlap* and *CoauCosine* than those who made their debut in earlier years. This means that young CS scholars start to work with specific coauthors but as they grow older academically, they keep finding new coauthors for conferences and journals (Figure 6 (a)) and diversify collaboration frequencies with coauthors who work together for both conference and journal papers (Figure 6 (b)). The per-debut-year *CoauOverlap* and *CoauCosine* trends appear mostly below 0.3 and 0.5, respectively, which are similar to the per-publication trend-lines in Figure 5, confirming the aforesaid observation that CS scholars seem to have distinct sets of coauthors for conferences and journals. A note is that while the *CoauOverlap* and *CoauCosine* in Figure 5 show steady increases overall as scholars publish more papers, those in Figure 6 decrease as scholars get older. This implies that older academic age (i.e., earlier debut year) is not strongly correlated with more papers in CS, which can be confirmed by the low Kendall's Tau (= 0.26) between length of academic age (2017 − debut year) and number of publication.

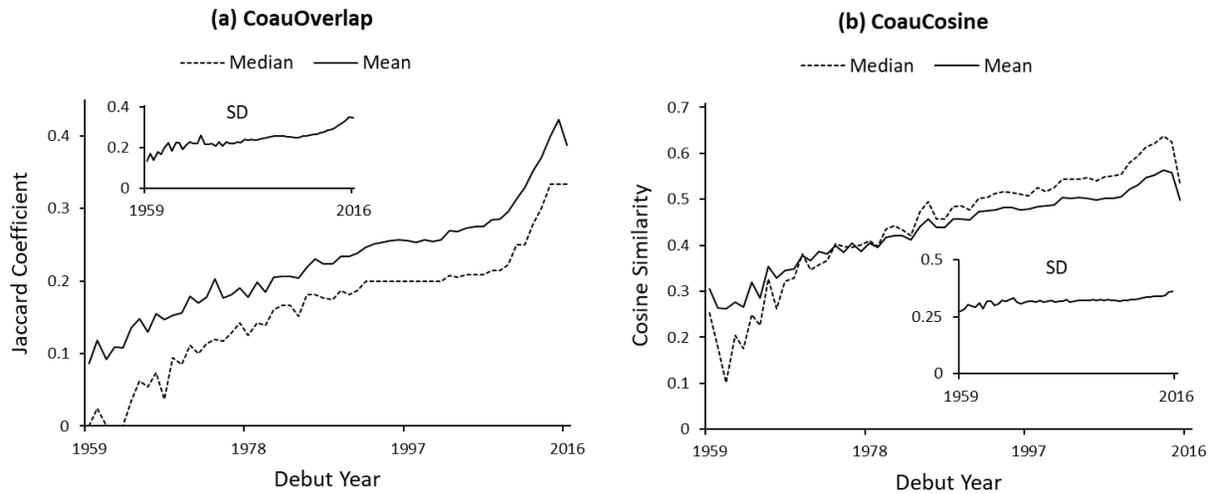

Figure 6: Trends of (a) Coauthor Overlap and (b) Cosine Similarity per Debut Year

*Title Words*

CS scholars in the data have used on average 41.49 unique title words in conferences and 30.16 unique title words in journals. However, the mean word count per paper is slightly lower for conferences (6.29) when compared with journals (6.95). This may indicate that journal titles tend to be more detailed or specific.

How do title words appear across conferences and journals? On average, unique title words in conferences and journals per author overlap for 16.79% (11.09 words) of all unique words per author (*WordOverlap* = 0.1679). The shared title words constitute on average 26.21% of all unique title words in conferences per author, while they constitute on average 34.57% in journals per author.

In Figure 7 (a), the *WordOverlap* ratio trend is shown over the number of publication. Despite a sharp value drop from two to three publications, the mean ratio of overlapping words between conference and journal papers per author tends to increase, although it hits the maximum below 0.3. If we assume that a conference paper is turned into a journal paper with exact or similar title words (e.g., Bar-Ilan, 2010), this observation implies that many CS conference papers may not be extended into journal papers (or vice versa). Previous studies found that the ratios of conference papers that transit into journals have been 25% ~ 33% in computer science (Bar-Ilan, 2010; Montesi & Owen, 2008; Wainer & Valle, 2013), which is comparable to the mean ratio of overlapping title words for conferences (26.21%) in this study. A note is that these preceding studies used papers as a unit of analysis, not individual authors. Despite such a difference in methodology, the author-based observations of this study seem to add evidence to findings that many conference papers are not turned into journal papers[7].

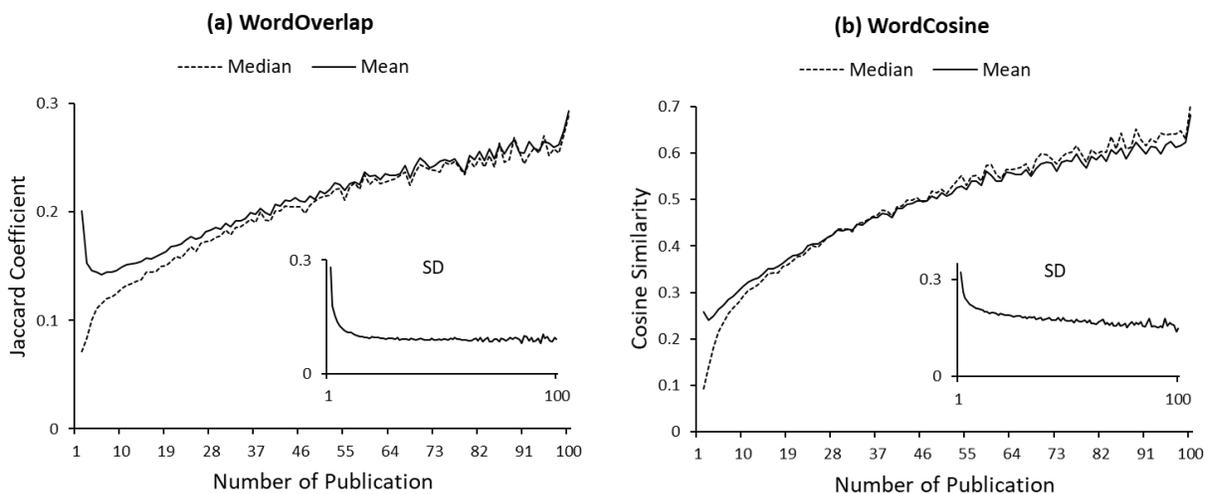

Figure 7: Trends of (a) Title Word Overlap and (b) Cosine Similarity per Number of Publication

Like *CoauOverlap*, the mean ratio of *WordOverlap* increases as authors publish more papers. A similar trend is also observed in Figure 7 (b), where the mean ratio of *WordCosine* is plotted over the number of publication. The trend starts at a very low level (below 0.3) but rises up close to 0.7. An implication of these increasing Jaccard coefficient and cosine similarity is that as CS scholars produce more papers, they come to focus on specific topics across journals and conferences, if we assume that title words can represent topics. This concentration of topics, however, seem to accompany topical diversity across conferences and journals, which is depicted by the rising but low mean ratios of overlapping words in Figure 7 (a).

In Figure 8, the *WordOverlap* and *WordCosine* ratio trend was plotted per debut year. Overall, authors who have published their first publication in recent years tend to show higher *WordOverlap* and *WordCosine* values than those who made their debut in earlier years. Especially, scholars whose debut years are between mid-1980s and mid-2010s show higher *WordCosine* values than other older and younger colleagues in Figure 8 (b). This means that CS scholars in this debut range work on more focused topics for both conferences and journals than others. But the per-debut-year *WordOverlap* and *WordCosine* trends appear mostly below 0.2 and around 0.3, respectively, indicating that many conference papers are not extended into journal work or vice versa. In addition, these trends contrast to increased trend-lines in Figure 7. Such difference may be mainly due to the fact that majority of CS scholars publish a small number of papers in both conferences and journals (see Figure 2). In Figure 7, the values of *WordOverlap* and *WordCosine* appear below 0.2 and around 0.3, respectively, for the production range between 2 and 17 which covers almost 80% of all authors in the data.

---

[7] This interpretation requires discretion. This study followed preceding studies in using title-word match as a proxy of measuring paper similarity via Jaccard Coefficient. In reality, authors may publish similar or same conference papers in journals (or vice versa) using different titles. Abstracts or full texts should be used to correctly capture the transition of conference papers into journal papers. However, this approach may not be feasible due to the difficulty in obtaining abstracts and full texts of all papers in the DBLP data used in this study.

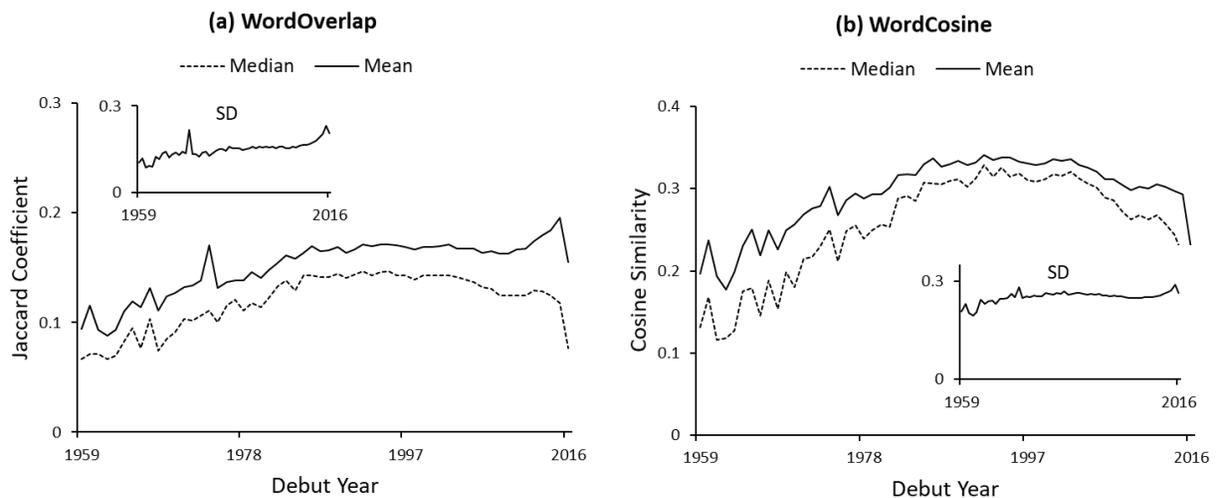

**Figure 8: Trends of (a) Title Word Overlap and (b) Cosine Similarity per Debut Year**

## Conclusion and Discussion

This paper identified publication patterns and trends in CS conferences and journals, using individual scholars as a unit of analysis. CS scholars were found to usually make their debut in conference papers, publish more papers in conferences than journals, and collaborate with more colleagues in conference publications than journal publications. CS scholars' focus on conference papers has begun around the 1980s, strengthened over time, and seems to continue in the near future. As such, conferences are the main vehicle of scholarly communication in CS.

An interesting finding is that overall conference publications do not seem to be preliminary work aimed at journal submission. This proposition is based on the observation that sets of coauthors per author across conferences and journals do not overlap much, which is also observed for sets of title words. Such observations may counter-argue the "worries that conference proceedings are merely a preceding step to a journal submission" that are often raised outside CS (Michels & Fu, 2014). According to the trends of conference-journal publication ratio, coauthor overlap, and title word overlap per author, CS conference publications look like having served as a distinct vehicle of research communication for several decades, corroborating findings of previous studies on this topic from an author-based approach.

These findings, however, do not imply that conferences should be prioritized over journals to understand computing research publications. Instead, a take-away of this study is that conference publication should be studied with special interests in order to properly understand CS scholars' scholarly communication. Bibliometrics studies have focused on journals as main outlets for disseminating research outcome and measuring performance of individual scholars. Such a journal-centric approach may be appropriate for other fields, but, for computing, it is not.

Some limitations apply. First, this study only considers the count of publication for representing research outputs. This approach ignores the content and quality of papers which are both important dimensions of scholarly impact. Also, this study does not consider the increase of conference venues, especially around the 1990s, when many conferences were established. This might contribute to the increased number of publications per CS authors. Second, the validity of all these findings is based on the coverage and correctness of the DBLP data. Although DBLP data have been assumed by many scholars to cover the majority of relevant publications in CS (Elmacioglu & Lee, 2005; Franceschet, 2011), its coverage is not perfect: i.e., its coverage of CS is different from other bibliometric databases such as IEEE Xplore, SCOPUS, and Web of Science (Reitz & Hoffmann, 2010; Wainer, Eckmann, Goldenstein, & Rocha, 2013). Also, some CS journals are indexed with some issues missing. For example, the first 22 volumes of *Journal of the Association for Information Science and Technology (JASIS&T)* are not included. Importantly, the accuracy of DBLP in identifying authors can be an issue. The DBLP name disambiguation showed a good performance against a sample of authors with most ambiguous names (Kim & Diesner, 2015). However, it surely has disambiguation errors due to faulty merging or splitting of unique identities, which may affect the outcomes of this study. Thus, the findings of this study should be understood to represent only the given dataset as it is.

Although this study showed that CS scholars tend to publish more in conferences than in journals, it could not give any clue to factors affecting such a propensity. Some possible factors can be listed as follows. Regarding publication cost, journals seem to be more attractive than conferences: publishing a paper in a journal is usually free of charge, but conferences generally require fees for accepted papers and authors to present to audience. However, conferences have a fixed time-table for publication (e.g., submission deadline and date of acceptance notification) and a lower time-to-market than journals (e.g., a few months from submission to acceptance or publication). In addition, legitimization of conference papers for a formal evaluation (Vardi, 2009) and quantitative evaluation might motivate CS scholars to submit more papers to conferences. For example, proceedings have been indexed in Scopus and Web of Science from 2004 and 2008, respectively (De Sutter & Van Den Oord, 2012) and surrogates to the *Journal Impact Factor*[8] have been used to rank conferences (e.g., the Australian CORE Ranking of Conferences). Furthermore, the opportunity to socialize with scholars and expose oneself to new research ideas and culture (esp., at international conferences in foreign countries) may incentivize CS scholars to prefer conferences to journals. A historical investigation into motivations of CS conference organizers in early days may reveal the origin and drive of burgeoning preference toward conferences.

As such, the topic of CS authorship in conferences and journals requires in-depth studies such as interviews to explore these factors besides data-driven analyses of publication patterns (Eckmann et al., 2012). Eventually, studying the conference versus journal differences in CS through various methodologies such as the author-level analysis of this study will help people outside CS better understand the unique publishing culture of CS and fairly evaluate conference and journal publication of CS scholars.

---

[8] https://clarivate.com/essays/impact-factor/

## Acknowledgements

I would like to thank Dr. Jana Diesner (University of Illinois at Urbana-Champaign) for comments on the initial versions of this manuscript. I am also thankful to anonymous reviewers for their helpful guidance on improving this paper.